\documentclass[aps,showpacs,onecolumn,floatfix,amsmath,amssymb,nofootinbib]{revtex4}
\usepackage{graphics}
\usepackage{epsfig}
\usepackage[english]{babel}
\usepackage{amsmath} 
\usepackage{verbatim} 
\usepackage{mathrsfs}
\usepackage{amsfonts}
\usepackage{amssymb}
\usepackage{soul} 
\usepackage[latin1]{inputenc}
\usepackage{hyperref}
\usepackage{cleveref}
\usepackage{float}
\usepackage{graphicx}
\usepackage{subfig}
\usepackage{color}
\usepackage{bm}

\newcommand{\lB}{\ensuremath{\ell_B}}
\newcommand{\beq}{\begin{equation}}
\newcommand{\eeq}{\end{equation}}
\newcommand{\barr}{\begin{eqnarray}}
\newcommand{\earr}{\end{eqnarray}}
\newcommand{\bea}{\begin{eqnarray*}}
\newcommand{\eea}{\end{eqnarray*}}

\DeclareMathOperator{\arcsinh}{arcsinh}

\begin{document}


\title{Magnetic contributions in Bekenstein type models}

\author{Lucila Kraiselburd$^{1,2}$} \email{lkrai@fcaglp.unlp.edu.ar}

\author{Florencia L. Castillo$^{3}$}
\email{casflo@ific.uv.es}

\author{Mercedes E. Mosquera$^{2,4,5}$}
\email{mmosquera@fcaglp.unlp.edu.ar}

\author{H\'{e}ctor Vucetich$^{1}$} \email{vucetich@fcaglp.unlp.edu.ar}

\affiliation{$^1$Grupo de Astrof\'{i}sica, Relatividad y Cosmolog\'{i}a, Facultad de Ciencias Astron\'omicas y Geof\'{\i}sicas, Universidad Nacional de La Plata, Paseo del Bosque S/N (1900) La Plata, Argentina \\
  $^2$CONICET \\
  $^3$Instituto de F\'isica Corpuscular, Consejo Superior de Investigaciones Cient\'{i}ficas y Universitat de Val\`{e}ncia,  Parque Cient\'{i}fico, C/Catedr\'{a}tico Jos\'{e} Beltr\'{a}n 2 (46980) Paterna, Espa\~{n}a.  \\
  $^4$ Facultad de Ciencias Astron\'omicas y Geof\'{\i}sicas, Universidad Nacional de La Plata, Paseo del Bosque S/N (1900) La Plata, Argentina.\\
  $^5$ Departamento de F\'isica, Universidad Nacional de La Plata,
  C.C. 67 (1900) La Plata, Argentina.}


\date{\today}


\begin{abstract}


  In this work we analyze the spatial and time variation of the fine
  structure constant ($\alpha$) upon the theoretical framework
  developed by Bekenstein \cite{Bekenstein02}. We have computed the
  field $\psi$ related to $\alpha$ at first order of the weak field
  approximation and have also improved the estimation of the nuclear
  magnetic energy and, therefore, their contributions to the source
  term in the equation of motion of $\psi$. We obtained that the
  results are similar to the ones published in Ref. \cite{LK11a} which
  were computed using the zero order of the
  approximation, showing that 
  one can neglect the first order contribution to the variation of the
  fine structure constant. By the comparison between our theoretical
  results and the observational data of the E\"otv\"os-type
  experiments or the time variation of $\alpha$ over cosmological
  time-scale, we set constrains on the free parameter of the
  Bekenstein model, namely the Bekenstein length.
\end{abstract}

\maketitle

\section{Introduction}
\label{sec:introduction}
Both, theories that have been developed to try to unify the
fundamental interactions (see, for example
\cite{wu,maeda,damour,brax,kaluza,wein}) as Dirac's large numbers
hypothesis (LNH) \cite{Dirac} predict that the fundamental constants
might vary in space time. Given these theoretical motivations, there
have been many attempts to look for observational and/or experimental
data to corroborate such variations (see Refs.
\cite{chand05,menegoni12,Fuji,peik} and references therein) but almost
all of them are until now consistent with a null variation. On the
other hand, the analysis of the data performed in
Refs. \cite{webb99,webb01,murphy01c,murphy03,King12} suggests that the
fundamental constants might have a different value than the present
ones. A space variation of fundamental constants would produce a
violation of the Weak Equivalence Principle (WEP) \cite{misner} which
can be verified thanks to the energy conservation \cite{hau}. The
E\"otv\"os-type experiments are the most sensitive methods to verify
the WEP since they are capable to detect the difference between the
acceleration of bodies with different composition or structure
\cite{Roll1964,Bragin1972,KeiFall82,Su94,Baessler:1999zz,Schlamm08}. For
details on the variation of the fundamental constants see \cite{uzan}.

Bekenstein proposed a theory to study the fine structure constant
variability based on first principles \cite{Bekenstein82} and later,
he claimed that no violations of WEP are perceived whereas a possible
time variation of $\alpha$ can not be discarded with the existed
observable bounds \cite{Bekenstein02}. Kraiselburd \& Vucetich
\cite{LK11a} modified Bekenstein's original theory by replacing the
classical model of charged particles by a quantum model of many bodies
at the non-relativistic limit, but with classical fields and
neglecting the contributions of the self-generated fields. They found
that even considering the cancellation proposed by Bekenstein, between
the electric field and the mass source term, quantum density
fluctuations produce fluctuating electric currents which generate
magnetic fields. These fields, are responsible for the appearance of
an anomalous term in the acceleration of a system of particles that
produces a violation of the WEP in E\"otv\"os-type experiments. Later,
Barrow \& Magueijo \cite{BM14,BM15} questioned Kraiselburd \&
Vucetich's interpretation of approaching the coupling of field $\psi$
in the expressions ${\bm D}=\epsilon{\bm E}=e^{-2\psi}{\bm E}$ and
${\bm H}=\mu^{-1}{\bm B}=e^{-2\psi}{\bm B}$ with the weak field
approximation ($e^{-2\psi}\approx 1$). 

The purpose of this work is twofold:
\begin{enumerate}
\item to show that nonlinear corrections to the zero order in Bekenstein's
  equations are negligible in the weak field regime, typical for
  experimental setups;
\item to analyze the prediction for the time variation of $\alpha$ due
  to the magnetic contributions in the semiclassical model and compare
  it with  modern observational data.
\end{enumerate}

 
This work is organized as follows. In Section \ref{sec:review} we
briefly review the formalism of Bekenstein model and Kraiselburd \&
Vucetich approach. In Section \ref{sec:nuevo} we introduce the
improvements to the semiclassical model in order to study the spatial
dependence, as well as the time variation of the fine structure
constant in an expanding Universe. The observational available data on
the variation of the $\alpha$ is shown in Section \ref{sec:data}. In
Section \ref{sec:results} we present the obtained bounds on the model
parameters through the comparison between the observational data and
the theoretical predictions. Finally in Section \ref{sec:conclusions}
we draw the conclusions.

\section{A review on Bekenstein's formalism and the semiclassical  approximation}
\label{sec:review}
In this section we summarize the main aspects of Bekenstein
model \cite{Bekenstein82,Bekenstein02} and its semiclassical version
presented in \cite{LK11a}.

\subsection{Bekenstein model}
\label{Bekmodel}
Bekenstein modifies Maxwell's electromagnetic theory to introduce a
scalar field $\psi$ that couples to charges generating a variation in
$\alpha$ \cite{Bekenstein82,Bekenstein02}. The model holds under the
following hypothesis
\begin{enumerate}
\item{Maxwell's theory must be recovered when $\alpha$ is constant;}
\item{the dynamic field $\psi$ is responsible for changes in
    $\alpha$;}
\item{the dynamics of the electromagnetic and the $\psi$ fields can be
    obtained from a variational principle};
\item{the theory must be gauge and time-reversal invariant, and also
    preserve causality;}
\item{it is not possible to have a length smaller than the Planck
    length $\ell_P$ \footnote {In string theories or in a low-energy
      limit of some TOEs (Theory of Everything), this statement should
      not be taken into account since, for these kind of theories,
      there are length scales that are smaller than the Planck
      scale. In these cases, this hypothesis should be revised.}.}
\end{enumerate}

From these hypothesis the only possible action is given by
\begin{eqnarray}
  \label{eq:Bek:Stot}
  S = S_{\rm em} + S_{\psi} + S_{\rm mat} + S_G,
\end{eqnarray}
where the expressions for the modified Maxwell and $\psi$ field
actions are
\begin{subequations}
  \label{eq:Bek:S_i}
  \begin{eqnarray}
    S_{\rm em} &=& -\frac{1}{16\pi}\int e^{-2\psi}f^{\mu\nu}f_{\mu\nu} \sqrt{-g} d^4x, \label{eq:Bek:Max}\\
    S_{\psi} &=& \frac{-\hbar c} {2\lB^2}  \int ( \partial_\mu\psi )^2 \sqrt{-g} d^4x, \label{eq:Bek:psi}      \\
    S_{\rm mat} &=&  \sum_i \int \frac{1}{\gamma}\left[-m c^2 +e \frac{u^{\mu}}{c} A_{\mu}\right] \delta^3\left[x^i - x^i (\tau)\right] d^4x, \label{eq:Bek:mat}\\
    S_G &=& \frac{c^4}{16 \pi G_N} \int \sqrt{(-g)} R d^4x, \label{eq:Bek:g}
  \end{eqnarray}
\end{subequations}
where $F_{\mu\nu}$ is the usual electromagnetic tensor,
$f_{\mu\nu} = e^{\psi}F_{\mu\nu}$. $g$ is the metric while $\hbar$ and
$c$ are the Planck constant and the speed of light respectively. $\lB$
is the {\it Bekenstein's fundamental length scale}, a constant length
introduced by the model. The Lorentz factor is represented by
$\gamma$, $m$ is the particle mass, $e$ and $u^\mu$ are its electric
charge and $4^{th}$ velocity respectively. Finally, $G_N$ is the
gravitational constant, $R$ the Ricci scalar, and $x^i (\tau)$ stands
for the particle trajectory of the particle in presence of the fields
$A_\mu$ and $\psi$.

Following the previous statements, an elementary electric charge and
the fine structure constant $\alpha$ are locally defined as
\begin{align}
  \label{eq:local:q}
  e(x^\mu) &= e_0 e^{\psi(x^\mu)},  & \alpha(x^\mu) &= e^{2{\psi(x^\mu)}} \alpha_0, 
\end{align}
where $e_0$, $\alpha_0$ are the
current values of the electron charge and the fine structure constant.

The equations that describe the dynamics of the electromagnetic and
$\psi$ fields are
\begin{subequations}\label{eq:Beq:em}
  \begin{equation}
    \left(e^{-\psi}F^{\mu\nu}\right)_{,\nu} = 4\pi j^\mu, \label{eq:Beq:Max:em}
  \end{equation}
  \begin{equation}
    \Box \psi = \frac{\lB^2}{\hbar c} \left(\frac{\partial\sigma}{\partial\psi}-\frac{F^{\mu\nu}F_{\mu\nu}}{8\pi} \right). \label{eq:Beq:psi:em}
  \end{equation}
\end{subequations}
In the previous equation $j^{\mu}=e_{0}u^{\mu}{\delta}^3[x^{i}-x^{i}(\tau)]\left(c\gamma\sqrt{-g}\right)^{-1/2}$,
$\sigma$ is the energy density of matter \cite{Bekenstein82}, and $\Box$
stands for the covariant flat d'Alambertian.

Bekenstein \cite{Bekenstein82} has analyzed the spatial
$\left(\nabla^2\psi\right)$ and the temporal
$\left(\frac{1}{c^2}\frac{\partial^2\psi}{\partial^2t}\right)$ terms
of Eq. (\ref{eq:Beq:psi:em}). The cosmological equation of motion for
$\psi$ is related to the time variation of $\alpha$. In an isotropic
and homogeneous Universe, with a Robertson-Walker metric, the time
variation of $\alpha$ (after calling $\epsilon=e^\psi$ and
$2 |\dot{\epsilon}/\epsilon|=|\dot{\alpha}/\alpha_0 |$) is given by
\begin{eqnarray}
  \frac{\partial}{\partial t}\left(a^{3}\frac{1}{\epsilon}\frac{\partial \epsilon}{dt}\right)=\frac{-a^{3}{\ell_B}^{2}}{\hbar c}\left[\epsilon\frac{\partial \sigma}{\partial \epsilon}+ \frac{(E^2-B^{2})}{4\pi} \right],\label{vartemp}
\end{eqnarray}
where $a$ represents the Universe scale factor. As one can see, the
only source term is the one that comes from matter, since the
electromagnetic radiation of the Universe does not
contribute. Considering that magnetic term can be neglected
such that the contribution of the Coulombic energy density is much higher
in ordinary matter and
$\epsilon\left(\frac{\partial\sigma}{\partial\epsilon}\right)\sim
2m_{EM}c^2$,
\begin{eqnarray}
  \frac{\partial}{\partial t}\left(a^{3}\frac{1}{\epsilon}\frac{\partial \epsilon}{dt}\right)=\frac{-a^{3}{\ell_B}^{2}}{c\hbar}\zeta_C\rho_mc^2,\label{vartemp2}
\end{eqnarray}
where $\rho_m$ is the total rest-mass density of electromagnetic
interacting matter $\left(m_{EM}\right)$ and $\zeta_C$ is the
fractional contribution of the Coulomb energy to the rest
mass. Mosquera et al. \cite{mosquera08} have analyzed
  the solution of the above equation for two different regimes of the
  Universe: when matter and radiation dominate, and when matter and
  cosmological constant do, matching both solutions with the
  appropriate boundary conditions. They have performed a statistical
  analysis to test the solutions using observational data from the
  early Universe and other bounds from the late Universe arriving to
  the conclusion that Bekenstein's hypothesis $\ell_B>\ell_P$ should
  be relaxed.

In Ref. \cite{Bekenstein82}, Bekenstein states that the mass dependent
term of Eq. (\ref{eq:Beq:psi:em}) is almost cancelled with the term of
the electric field, therefore
$\frac{\partial\sigma}{\partial\psi}+\frac{e^{-2\psi}E^2}{4\pi}\approx
0$.
He justifies this statement with the fact that field $\psi$ can be
written as a function of the coulombic potential $\phi_C$, and
working on this relationship, he derives that there exists an
asymptotically a null value of $\psi$ \cite{Bekenstein02}\footnote{If this statement is valid the restrictions found in Ref. \cite{mosquera08} cease to make sense since $\frac{\partial}{\partial t}(a^{3}\frac{1}{\epsilon}\frac{\partial\epsilon}{\partial t})=0$ considering only the Coulomb contribution.}. Hence, an
electrically charged particle (or a set of point charges) satisfies
$\Box \psi\approx 0$, meanwhile for a static system of magnetic
dipoles,
$\nabla^2 \psi\approx \frac{-\lB^2}{\hbar
  c}e^{-2\psi}\frac{B^2}{4\pi}$
and
$\frac{\partial}{\partial t}(a^{3}\frac{1}{\epsilon}\frac{\partial
  \epsilon}{\partial
  t})=\frac{a^{3}{\ell_B}^{2}}{c\hbar}|\zeta_m|\rho_mc^2$.
The factor $\zeta_m$ is the fractional contribution of the magnetic
energy to the rest mass. For this latter system, Bekenstein claims
that no WEP violation can be detected in laboratory experiments.
 
\subsection{The semiclassical approach}
\label{KVAP}
In Bekenstein's theory particles are treated classically. This
assumption is not accurate, especially in those cases where quantum
effects become relevant, such as white dwarfs or condensed matter
physics, or at high-energy scales (a small distance scales) since
fermions have their own ``natural length scale''. For these reasons,
Kraiselburd \& Vucetich \cite{LK11a} have proposed a semiclassical
model where particles are quantized and fields remain classic. Once
matter is quantized, the main subject is to understand matter
distributions movements inside a scalar field $\psi$, and the
contributions of the electric and magnetic field of this interaction.

The full action and Lagrangian of a body (or an ensemble of point
charges) have been analyzed in the non-relativistic limit for the
charges, with the full expression for the electromagnetic field. If
only the external $\psi$ and gravitational fields are taken into
account, meaning that the self-generated fields are neglected, the
equation of motion for the scalar field $\psi$ is
\begin{eqnarray}
  \label{LKCap2}
  \Box\psi=4\pi\kappa^2\left[\frac{\partial(\bar{\Psi}\partial m\Psi)}{\partial\psi}+\frac{F_{\mu \nu} F^{\mu\nu}}{8\pi}\right],
\end{eqnarray}
being $\kappa^2=\frac{\ell_B^2}{4\pi\hbar c}$. $\Psi$ represents the
charged particle field, while the particle mass has been renormalized
as $m=m_0+\partial m$ with $m_0$ independent of $\psi$. The expression
$\partial m$ can only be computed through the quantum electrodynamics
renormalization since it arises from the quantum fluctuations of the
electromagnetic field,
$\partial m\approx\frac{3}{4\pi}e^{2\psi} \alpha_0 m_0
\log\left(\frac{\Lambda^2}{m_0^2}\right)$,
with $\Lambda$ the ultraviolet cut-off frequency imposed by quantum
electrodynamics. Under this scheme, it is not possible to determine if
Bekenstein's cancellation statement (in this case,
$\frac{\partial(\bar{\Psi}\partial m\Psi)}{\partial
  \psi}+e^{-2\psi}\frac{E^2}{4\pi}\approx 0$)
is valid or not since it is necessary to quantify the fields
\cite{LK12}.

Even taking Bekenstein's statement as valid, in a quantum model of
matter, quantum fluctuations of protons and neutrons oscillating in
anti-phase create a dipole moment which produces a
variable current and  magnetic field. Haugan and Will
\cite{Haugan77,Will81} have made estimations of these contributions to
the magnetic energy from a minimal nuclear shell model roughly
obtaining
\begin{eqnarray}
  \label{eq:Bek:Mag:Ener}
  E_m \simeq \frac{3}{20\pi} \frac{\hat{E}}{R(A)\hbar c} \int\sigma dE = \frac{0.24}{\pi} \frac{\hat{E}}{R(A)\hbar c},
\end{eqnarray}
where $R(A)=1.2 A^{1/3}{\rm fm}$ and $\hat{E}\sim 25 {\rm MeV}$ are
the nuclear radius and the giant dipole mean absorption energy
respectively, and $A$ is the atomic
number. The integrated strength function $\int\sigma dE$ satisfies the Thomas-Reiche-Kuhn sum rule  but for light nuclei, this approximation is no longer suitable. Consequently, in this article we use other estimates for nitrogen, hydrogen and beryllium (see Appendix \ref{NME}).\\

The magnetic energy density $e_m$, which is concentrated near atomic
nucleus, can be written in terms of the fractional contribution of the
magnetic energy to the rest mass of nuclear specie $b$
$\left(M_b\right)$, $\zeta_m^b = \frac{E_m^b}{M_bc^2}$ as,
\begin{eqnarray}
  \label{eq:Bek:Mag:Eden}
  e_m(\bm{x}) = \sum_a E_m^a \delta(\bm{x} - \bm{x}_a) \simeq
  \sum_b{E}_m^b n_b(\bm{x})\simeq \frac{\sum_b \zeta_m^b \rho_b(\bm{x})}{\rho(\bm{x})}\rho(\bm{x})c^2= \bar{\zeta}_m(\bm{x})\rho(\bm{x})c^2, 
\end{eqnarray} 
where
$\bar{\zeta}_m(\bm{x}) = \frac{\sum_b \zeta_m^b
  \rho_b(\bm{x})}{\rho(\bm{x})}$ is
the local mass-weighted average of $\zeta_m$ and $n_b(\bm{x})$ is the 
particle density of the nuclear specie $b$. Finally, $\rho_b(\bm{x})$
stands for the mass density of the nuclear specie $b$, and
$\rho(\bm{x})$ is the local mass density.

Ref. \cite{LK11a} is based specifically on the analysis of the
magnetic contributions in the spatial part of a semiclassical
Bekenstein model. Taking the spatial part of Eq. (\ref{LKCap2}) and
rewriting it using expression of Eq. (\ref{eq:Bek:Mag:Eden}) it
becomes
\begin{eqnarray}
  \label{eq:Bek:em:psi:rho}
  \nabla^2\psi = -8\pi\kappa^2c^2 e^{-2\psi}\bar{\zeta}_m \rho.
\end{eqnarray}
A   solution for an arbitrary distribution of sources with spherical symmetry of the scalar
field $\psi$ under the weak field approximation ($e^{-2\psi}\sim 1$)
can be found as,
\begin{eqnarray}
  \label{eq:Bek:Sol:psi}
  \psi(r) = 8\pi\kappa^2c^2  \int_0^{R_s} \frac{x^2}{r_>} \bar{\zeta}_m(x) \rho(x) dx,
\end{eqnarray}
being $R_s$ the distribution radius and
\begin{equation}
  r_>=\begin{cases} r &\mbox{if } r>x \\ x &\mbox{if } x>r\end{cases}\nonumber
\end{equation}
Then, using the Newtonian potential, $\phi_N(r)=G_N M/r$, the field outside the distribution ($r>x$) can be written as
\begin{eqnarray}
  \label{eq:Bek:Sol:psi:fi}
  \psi(r) \asymp  \frac{8\pi c^2\kappa^2}{G_N M} \phi_N(r) \int_0^{R_s} x^2
  \bar{\zeta}_m(x)   \rho(x) dx = 
  2 \left(\frac{\ell_B}{\ell_P} \right)^2 \tilde{\zeta}_m
  \frac{\phi_N(r)}{c^2}.
\end{eqnarray}

In the last expression, $\tilde{\zeta}_m$ is the mass-averaged value
of $\bar{\zeta_m}$. The Lagrangian of the theory, according to
Ref. \cite{LK11a}, is
\begin{eqnarray}
  \label{17paper}
  L = - M_{tot} \left[ c^2 - \frac{V_{CM}^2}{2} - \phi_N(\bm{R}_{CM}) \right] + 2 \psi(\bm{R}_{CM}) E_m + \cdots,    
\end{eqnarray}
where $M_{tot}$ is the body mass, $V_{CM}$ is the velocity of the
center of mass and $\bm{R}_{CM}$ is the center of mass
position. Recalling the assumption of the cancellation between the
electrostatic contribution and mass dependence on $\psi$, one can
write the acceleration that suffers a body immersed in an external
gravitational $\left(\bm{g}\right)$ and $\psi$ fields as
\begin{eqnarray}
  \label{18paper}
  \ddot{\bm{R}}_{CM} = \bm{a }= \bm{g}+2 \frac{E_m}{M_{tot}} \nabla\psi \left|_{CM} \right.,
\end{eqnarray}
with $\nabla\phi_N(r)={\bm g}$. The neglected terms
in the last expression are of either tidal order, negligible in
laboratory tests of WEP, or of higher order in $\psi$. The last term
on the right-hand side of the above equation is an anomalous
acceleration generated by the scalar field $\psi$. Consequently, the
validity of the theory can then be tested with E\"otv\"os-type
experiments that measure the correlation between inertial mass and
gravitational mass. From expressions of Eqs. (\ref{eq:Bek:Sol:psi:fi})
and (\ref{18paper}) it is possible to compute the E\"otv\"os parameter
which associated with the differential acceleration of two bodies ($A$
and $B$) of different composition
\begin{eqnarray}
  \label{eq:Bek:eta:expr}
  \eta(A, B) = \frac{a_A - a_B}{g} =4 \left(\frac{\ell_B}{\ell_P}\right)^2 \zeta^S_m (\zeta^{A}_m - \zeta^{B}_m),
\end{eqnarray}
where $\zeta^S_m$, $\zeta^{A}_m$ and $\zeta^{B}_m$
are the magnetic energy fractions of the source and
bodies $A$ and $B$ respectively.

The bounds on $\left(\frac{\ell_B}{\ell_P}\right)^2$ obtained in
Ref. \cite{LK11a} using the data of the most accurate versions of the
E\"otv\"os experiment presented in Table \ref{tab:EotRes}, are
\begin{eqnarray}
  \label{eq:Res1:lBlP2}
  \left(\frac{\ell_B}{\ell_P}\right)^2 = 0.0003 \pm 0.0006,  \hskip 1cm  
  \frac{\ell_B}{\ell_P} &< 0.05, 
\end{eqnarray}
where the last equation correspond to the upper bound at $3\sigma$ level.

From these results it is clear that even when the electric field does
not contribute to the scalar field $\psi$, the quantum fluctuations
contribute (through currents) in such a way that the upper limit
found for the $\ell_B/\ell_P$ parameter discards the theory.

\begin{table}[H]
  \centering
  \begin{tabular}{cccr@{$\pm$}lr}
    \hline
{\bf Body $A$ } & {\bf Body $B$ } & {\bf Source} & \multicolumn{2}{c}{{\bf $10^{11}\eta(A,B)$}} & {\bf Ref.}\\  
    \hline
    \hline
    Al & Au & Sun & 1.0 & 1.5 & \cite{Roll1964}\\
    Al & Pt & Sun & 0.03 & 0.045 & \cite{Bragin1972}\\
    Cu & W  & Sun & 0.0 & 2.0 & \cite{KeiFall82}\\
    Be & Al & Earth & -0.02 & 0.23 & \cite{Su94}\\
    Be & Cu & Earth & -0.19 & 0.25 & \cite{Su94}\\
    Be & Al & Sun & 0.40 & 0.98 & \cite{Su94}\\
    Be & Cu & Sun & -0.51 & 0.61 & \cite{Su94}\\
    Si/Al & Cu & Sun & 0.51 & 0.67  & \cite{Su94}\\
    EC & MM & Sun & 0.001 & 0.032 & \cite{Baessler:1999zz}\\
    Be & Ti & Earth & 0.004& 0.018 & \cite{Schlamm08}\\
    \hline
  \end{tabular}
  \caption{E\"otv\"os-type experiments data.}
  \label{tab:EotRes}
\end{table}

\section{Higher orders in the weak field approximation under the
  semiclassical approach}
\label{sec:nuevo}
In this section we present the computation of the field $\psi$ by
taking into account the following order in the series expansion in the
equation of motion under the scheme proposed by Kraiselburd \&
Vucetich \cite{LK11a}.

In order to carry out this calculation, we proceed to solve
Eq. (\ref{eq:Bek:em:psi:rho}) with an iterative method
inspired in the Picard method of solution for
  differential equations. Considering a weak field, $e^{-2\psi}\approx1-2\psi+ O( \psi^{2})$, we begin by
introducing this approximation of the exponential in
the equation of motion (see Eq. (\ref{eq:Bek:em:psi:rho})). Then,
we replace the scalar field
$\psi$  on its right-hand side (source term) with the solution given by Eq. (\ref{eq:Bek:Sol:psi:fi}), that
is the zero order approximation $\psi_0(r)$, to find the next order
solution $\psi_1(r)$,
\begin{eqnarray}
  \nabla^2\psi_1 &=& -8\pi\kappa^2c^2  \bar{\zeta}_m \rho(1-2\psi_0), \nonumber\\
                 &=& -8\pi\kappa^2c^2\bar{\zeta}_m \rho \left[1-4\left(\frac{\ell_B}{\ell_P} \right)^2 \tilde{\zeta}_m \frac{\phi_N(r)}{c^2}\right],
                     \label{psiPicard}
\end{eqnarray}
The solution is then
\begin{eqnarray}
  \psi_1(r) &=& 8\pi\kappa^2c^2 \frac{1}{r} \int_0^{R} x^2 \bar{\zeta}_m(x)\left[1-4\left(\frac{\ell_B}{\ell_P} \right)^2 \tilde{\zeta}_m\frac{\phi_N(x)}{c^2} \right]\rho(x) dx ,\label{psi1L} \nonumber\\
            &=&\psi_0(r) -\frac{32\pi\kappa^2}{G_{N} M} \phi_N(r)\left(\frac{\ell_B}{\ell_P} \right)^2\tilde{\zeta}_m\int_0^R x^2  \bar{\zeta}_m(x) \phi_N(x) \rho(x) dx,  \label{psi1C}
\end{eqnarray} 
where $R$ is the radius and $M$ is the mass, both of the source
body. The previous integral can be computed as
\begin{eqnarray}
  \label{psiinte}
  \int_0^R x^2 \bar{\zeta}_m(x) \phi_N(x) \rho(x) dx = \frac{3}{2}\phi_N(R) \tilde{\zeta}_m M,
\end{eqnarray}
where, once again, $\tilde{\zeta}_m$ stands for the mass average of
$\bar{\zeta}_m$. Therefore, the asymptotic behaviour of the field
turns out to be
\begin{eqnarray}
  \label{psi1asin}
  \psi_1(r) =\left[1 - 6 \left(\frac{\ell_B}{\ell_P} \right)^2 \tilde{\zeta}_m\frac{\phi_N(R)}{c^2}\right] \psi_o(r).
\end{eqnarray} 

For E\"otv\"os-type experiments in which generally the source is the
Earth or Sun, the factor $6 \tilde{\zeta}_m\frac{\phi_N(R)}{c^2}$ of
the second term inside the bracket is of the order
$\sim 2.37\times 10^{-10}$ and $\sim 5.27\times10^{-14}$
respectively. Therefore, the contribution of this term would be
negligible.

\subsection{Magnetic contributions to the acceleration}
\label{magespacial}

In this section we compute and discuss the acceleration changes due to
the addition of the new computed field $\psi_1$. According to
Eq. (\ref{psi1asin})
\begin{eqnarray} \label{Gragpsi1}
  \nabla\psi_1 |_{{\rm CM}} =
      2\left(\frac{\ell_B}{\ell_P} \right)^2 \zeta_{m}\frac{\nabla\phi_N(r)}{c^2}-  12\left(\frac{\ell_B}{\ell_P} \right)^4 \zeta_{m}^2\frac{\phi_N(R)\nabla\phi_{N}(r)}{c^4}.
\end{eqnarray}
Consequently, the acceleration that a body $j$ in presence of a with
the Newtonian gravitational and a scalar fields becomes (see
Eq. (\ref{18paper}))
\begin{eqnarray}
  {\bm a}_j={\bm g}+4\zeta^{j}_m\zeta^S_{m}\left(\frac{\ell_B}{\ell_P} \right)^2{\bm g}-24\zeta^{j}_m{\zeta^S_{m}}^2\left(\frac{\ell_B}{\ell_P} \right)^4\frac{\phi_N(R)}{c^2}{\bm g}\label{acbsc}.
\end{eqnarray}

One can observe again an anomaly, which is composed of two terms. The
first of them agrees with the one described in the previous section
(see Eq. (\ref{eq:Bek:eta:expr})), while the second one is the result
of the next order of the field $\psi$ expansion described in this work.

\subsection{Magnetic contributions to the time variation of $\alpha$
  in an expanding Universe}
\label{magtemporal}

In this section we analyze the time variation of the fine structure
constant using the development carried out in Ref. \cite{mosquera08}
for an expanding Universe. This calculation is performed by taking
into account the magnetic contribution of the baryon matter.

From Eq. (\ref{vartemp}), considering valid Bekenstein's cancellation
in the semiclassical approach, we
obtain \begin{equation}\label{vartempep}
  \frac{1}{\epsilon}\frac{\partial \epsilon}{\partial t}=
  \frac{3}{4\pi}\left(\frac{\ell_B}{\ell_P}\right)^2\zeta_m\Omega_b\left(\frac{a_0}{a(t)}\right)^3H^2_0
  (t-t_c),
\end{equation}
where we have used $\rho_m=\frac{\Omega_b\rho_C}{a^3(t)}$,
$\rho_C=3/(8\pi G_N)H_0^2$ and $\ell_P=(G_N\hbar/c^3)^{1/2}$, $H_0$ is
the Hubble constant, $\Omega_b$ is the Universe baryon matter density
and $t_c$ is an integration constant that can be null. One can assume
that the field variation is only generated by the magnetic energy of
the cosmological baryon matter. If one consider that the primordial
Universe is composed only of Hydrogen and Helium-4, the fractional
contribution of the magnetic energy to the rest baryon mass is given
by
\begin{equation}\label{zetamag}
  \zeta_m=\frac{ 0.75 E_m^H + 0.25 E_m^{^4He}}{0.75 M_{\rm H}c^2  + 0.25 M_{^4{\rm He}}c^2}\approx 1.097\times10^{-5},
\end{equation}
where $E_m^H$ and $E_m^{^4He}$ are the magnetic energy of H and $^4$He respectively and $M_{\rm H}$ $\left(M_{^4{\rm He}}\right)$ is the H ($^4$He) mass. These magnetic energies in particular are estimated differently from the others (an explanation is given in appendix \ref{NME}).\\

The scale factor has different expressions depending on the stage of
the Universe evolution. In a flat Friedmann-Robertson-Walker metric
(FRW) and assuming the scalar field contribution null, the equation
that describes the Universe scale factor time evolution can be written
as
\begin{equation}\label{24Mer}
  \left(\frac{1}{a}\frac{\partial a}{\partial t}\right)^2 =H_0^2\left\{\Omega_m \left(\frac{a_0}{a(t)}\right)^3 + \Omega_r \left(\frac{a_0}{a(t)}\right)^4 + \Omega_{\Lambda}\right\},
\end{equation}
where the initial conditions are $a(0)=0$ and $a(t_0)=1$, being $t_0$
the age of the Universe, $\Omega_m$ is the matter contribution to the
density, $\Omega_r$ is the radiation density and $\Omega_{\Lambda}$
stands for the contribution of the cosmological constant to the
density. In Ref. \cite{mosquera08}, the authors have performed a
piece-wise approximate solution by solving the FRW equation for two
different cases: a) radiation and matter (early Universe) that can be
applied to nucleosynthesis and recombination of primordial hydrogen;
and b) matter and cosmological constant (middle-age and present
Universe) which is used for quasar absorption systems, geophysical
data, and atomic clocks. Then, they have matched both solutions in the
time that the regime changes $\left(t_1\right)$ and considered that
the scale factor is a smooth and continuous function of time. They
have obtained
\begin{itemize}
\item early Universe, dominated by radiation and matter
  $\left(t<t_1\right)$
  \begin{subequations}
    \begin{eqnarray}
      a_{RM}(\xi) &=& \frac{\Omega_m \xi^2}{4} + \sqrt{\Omega_r}\xi,\label{25Mer}\\
      H_0 t(\xi)&=&\frac{\Omega_m \xi^3}{12} +\frac{ \sqrt{\Omega_r} \xi^2}{2}.\label{26Mer}
    \end{eqnarray}
  \end{subequations}
\item middle-age Universe, dominated by matter and cosmological
  constant $\left(t>t_1\right)$
  \begin{equation}\label{27Mer}
    a_{MC} = \sqrt[3]{\frac{\Omega_m}{\Omega_{\Lambda}}} \left[ \sinh\left( \frac{3}{2} \sqrt{\Omega_{\Lambda}} H_0 (t-t_0) + \arcsinh \left(\sqrt{ \frac{\Omega_{\Lambda}}{\Omega_m}}\right)\right) \right]^{\frac{2}{3}}. 
  \end{equation}
\end{itemize}

Thereafter, we obtain the time variation of $\alpha$ (being
$\ln
\frac{\epsilon(t)}{\epsilon(t_0)}\simeq\frac{1}{2}\frac{\Delta\alpha}{\alpha_0}$)
for each stage of the Universe by replacing the scale factor in
Eq. (\ref{vartempep});
\begin{enumerate}
\item early Universe dominated by radiation and matter
  $\left(t<t_1\right)$
  \begin{subequations}
    \begin{eqnarray}
      \frac{\Delta\alpha}{\alpha_0} &=& \frac{-1}{\pi} \zeta_m \frac{\Omega_b}{\Omega_m}  \left(\frac{\ell_B}{\ell_P}\right)^2\left[\ln\left(  \frac{ \lambda(\xi_1)}{ \lambda(\xi)}\right) + \frac{2 \sqrt{ \Omega_r}}{ \lambda(\xi)} - \frac{2 \sqrt{ \Omega_r}}{ \lambda(\xi_1)} - \frac{1}{8} \ln\left( \frac{ \Omega_r}{ \Omega_{\Lambda}}\right)\right] \nonumber \\
                                    && + \frac{3}{4\pi} \zeta_m  \left(\frac{\ell_B}{\ell_P}\right)^2 \frac{\Omega_b}{\Omega_r}H_0t_c\left[\frac{1}{\xi} - \frac{1}{\xi_1} + \frac{\Omega_m}{ \lambda(\xi)} - \frac{\Omega_m}{\lambda(\xi_1)} + \frac{\Omega_m}{2\sqrt{\Omega_r}}\ln\left(\frac{ \xi  \lambda(\xi_1)}{ \xi_1  \lambda(\xi)}\right)\right]\nonumber \\
                                    &&-\frac{1}{2\pi} \zeta_m \frac{\Omega_b}{\Omega_m} \left(\frac{\ell_B}{\ell_P}\right)^2 \sqrt{\Omega_{\Lambda}}\left[ \frac{\tau_1 - H_0t_c}{ \tanh\left(\arcsinh\left(\left(\frac{\Omega_r \Omega_{\Lambda}}{\Omega_m^4}\right)^{\frac{3}{8}} \right)\right)} - \frac{\tau_0 - H_0t_c}{ \tanh \left( \arcsinh \left( \sqrt{ \frac{ \Omega_{\Lambda}}{\Omega_m}}\right)\right)}\right], \label{var_al_ut} \\
      \lambda(\xi)&=&\xi \Omega_m + 4 \sqrt{\Omega_r}, \label{lam}\\
      \tau_i &=&H_0t_i.
    \end{eqnarray}
  \end{subequations}

\item middle-age Universe dominated by matter and cosmological
  constant $\left(t>t_1\right)$
  \begin{subequations}
    \begin{eqnarray}
      \frac{\Delta\alpha}{\alpha_0} &=& \frac{-1}{2\pi} \zeta_m \frac{\Omega_b}{\Omega_m} \left(\frac{\ell_B}{\ell_P}\right)^2\sqrt{\Omega_{\Lambda}}\nonumber \\ 
                                    &&\times \left[ \frac{ \tau - H_0 t_c}{\tanh \left(\gamma (\tau)\right)} - \frac{\tau_0 - H_0t_c}{\tanh \left(\arcsinh \left({ \sqrt{ \frac{ \Omega_{\Lambda}}{ \Omega_m}}}\right)\right)} - \frac{2}{3 \sqrt {\Omega_{\Lambda}}} \ln\left( \sqrt{ \frac{ \Omega_m}{ \Omega_{\Lambda}}} \sinh \left(\gamma(\tau)\right)\right)\right],\label{var_al_um} \\
      \gamma(\tau) &=&\frac{3}{2} \sqrt{\Omega_{\Lambda}} (\tau - \tau_0) + \arcsinh\left(\sqrt{\frac{\Omega_{\Lambda}}{\Omega_m}}\right). \label{gam}
    \end{eqnarray}
  \end{subequations}

\item present Universe dominated by matter and cosmological constant
  $\left(t=t_0\right)$
  \begin{equation}
    \label{var_al_ua}
    \frac{\dot\alpha}{H_0\alpha_0}= \frac{3}{4\pi}\zeta_m\left(\frac{\ell_B}{\ell_P}\right)^2 \Omega_b (\tau_0-\tau_c).
  \end{equation}
\end{enumerate}

\section{Observational data on the time variation of the fine
  structure constant}
\label{sec:data}
In this section we present the observational data on the time
variation of the fine structure constant. To study this model of the
time variation of $\alpha$, we take into account not only the
different stages in the Universe's evolution but also use three
different sets of cosmological parameters
\cite{menegoni12,planck15}. Since the baryon density
($\Omega_b\, h^2$) can be determined by two independent analysis: i)
by the comparison between theoretical calculation of primordial
abundances and the observable data; or ii) by the study of the CMB
data (Cosmic Microwave Background)
\citep{planck16,planck14,wmap13,wmpa09}, we used the results of the
analysis of the CMB data. That is
$\Omega_b\, h^2=0.02212 \pm 0.00028$,
$\Omega_b\, h^2=0.0231 \pm 0.0013$ \citep{planck15},
$\Omega_b\, h^2=0.0218 \pm 0.0004$ \cite{menegoni12}.

\subsection{BBN constraints}
\label{sec:BBN}
During the first three minutes of the Universe the light elements,
such as deuterium, $^4$He, $^3$He, $^7$Li, were produced. 
In order to include the fine structure constant time variation upon
cosmological time-scale, we have modified the numerical code
developed by Kawano et al. \cite{kawano88,kawano92} (for the
modifications to the code see
\cite{nollet02,iguri99,landau06,landau08,mosquera08} and references
therein). As said before, the baryon density was fixed at three
different values.

The possible variation of $\alpha$ has been calculated through the
comparison of the observable data and the theoretical results using a
$\chi^2$-test. For the deuterium, we have considered the observable
data reported in
Refs. \cite{pettini01,omeara01,kirkam03,omeara06,oliveira06,crighton04,cooke16,cooke14,pettini12}. For
$^4$He we used the data of
Refs. \cite{izotov07,izotov04,izotov98,thuan02,villanova09,izotov13,aver13,izotov14,luridiana03},
and the results of
Refs. \cite{bonifacio02,boesgaard05,molaro97,melendez10,mucciarelli14,nissen12,lind09}
for $^7$Li.

In order to check the consistency of the data we have followed the
analysis of Patrignani et al. \cite{pdg16} and increased the
observational errors by a factor $\Theta_{\rm D}=1.28$,
$\Theta_{^4\rm He}=2.79$ and $\Theta_{^7\rm Li}=2.04$. In Table
\ref{BBN} we present the results. The inclusion of the lithium data
gives a rather poor fit for all the models, meaning for the three
values of $\Omega_b h^2$. If the middle value of the baryon density is
used (Model II), the fit is excellent and the variation of $\alpha$ is
consistent with a null-variation at $1\sigma$, if one excludes the
$^7$Li data in the analysis. The fits obtained with the largest (Model
III) or the smallest (Model I) value for $\Omega_B \, h^2$ considered
in this work, are good fits, however, $\frac{\Delta \alpha}{\alpha}$
is consistent with zero at $2\sigma$ if the lithium data is removed in
the statistical test.  \renewcommand{\arraystretch}{1.7}
\begin{table}[H]
  \begin{center}
    \begin{tabular}{ccccc}
      \hline
      {\bf Model}&{\bf Data}&$\Omega_b\, h^2$ &$\frac{\Delta \alpha}{\alpha} \pm \sigma$ $[10^{-2}]$ & $\chi^2_{\nu}/(N-1)$ \\ \hline\hline
      Model I& All data &$0.0218 \pm 0.0004$&$0.24^{+0.30}_{-0.28}$&$3.55$\\
                 &Without $^7$Li-data& &$-0.32^{+0.30}_{-0.32}$&$1.24$\\\hline

      Model II& All data &$0.02212 \pm 0.00028$&$0.52^{+0.33}_{-0.26}$&$3.54$\\
                 &Without $^7$Li-data&  &$-0.05^{+0.26}_{-0.35}$&$1.01$\\\hline

      Model III& All data &$0.0231 \pm 0.0013$&$1.42^{+0.22}_{-0.32}$&$4.20$\\
                 &Without $^7$Li-data& &$0.65^{+0.26}_{-0.35}$&$1.25$\\\hline
    \end{tabular}
    \caption{Time variation of the fine structure constant for the
      different values of the baryon density and considering all data
      in the sample or removing the $^7$Li data in the $\chi^2$-test.}
    \label{BBN}
  \end{center}
\end{table}

\subsection{CMB constraints}
The recombination process, when the first hydrogen atoms formed and
the Universe became transparent to photons, occurred approximately
$300000$ years after the Big Bang. The photons released during this
process are the Cosmic Microwave Background radiation (CMB). This
radiation corresponds to a black body radiation at temperature
$T_0 = 2.725 \, ^0{\rm K}$ with small anisotropies of the order of
${10^{-6}} \,^0{\rm K}$.

The CMB formation is totally mediated by electromagnetic
processes. Thus, a variation in the coupling constant of this
interaction (that is $\alpha$) involves modifications in the
interactions between photons and electrons as well as in the
recombination scenario and consequently in the spectrum of CMB
fluctuations \cite{Landau02,menegoni12,planck15}. In Table
\ref{tablaUniTemp} we show the bounds obtained in
Ref. \cite{menegoni12,planck15} using WMAP7, WMAP9 and Planck
satellites data.  \renewcommand{\arraystretch}{1.7}
\begin{table}[H]
  \begin{center}
    \begin{tabular}{ccccc}
      \hline
      {\bf Model} &{\bf Data source}&{\bf $\frac{\Delta\alpha}{\alpha_0}\pm\sigma $}&{\bf Ref.}\\ \hline\hline
      Model I & WMAP7 & $ -0.016\pm 0.005$&\cite{menegoni12}\\
      Model II & Planck & $-0.0036\pm0.0037$& \citep{planck15}\\ 
      Model III & WMAP9 & $ 0.007\pm0.020$&\cite{planck15}\\\hline
    \end{tabular}
    \caption{Bounds on the time variation of $\alpha$ from CMB data}
    \label{tablaUniTemp}
  \end{center}
\end{table}

\subsection{Quasars absorption spectra}
Spectroscopic observations of extragalactic objects such as quasars
can be used to study the time variation of the fine structure
constant. High redshift quasar spectra present absorption resonant
lines of alkaline ions whose splits are proportional to
$\alpha^2$. Therefore, studying the separation of the doublets, one can
obtained bounds for the variation of $\alpha$. This analysis (called
AD) have been applied, in the literature, to different doublet
absorption lines systems at different redshifts
\cite{cowie95,murphy01b,chand05}. The results are consistent with a
null variation of the constant.

The Many Multiplet Method (MMM) which compares transitions of
different species with widely differing atomic masses together with
different transitions of the same species, improves the sensitivity,
but its systematic errors are generating great controversies
\footnote{From the combined observations of distant quasars using the
  KECK/HIRES \cite{webb99,webb01,murphy01c,murphy03} and VLT/UVES
  \cite{King12} telescopes, there seems to exist substantial evidence
  for a dipole-type spatial variation of $\alpha$. Pinho \& Martins
  \cite{Pinho} have reached the same conclusion from an independent
  analysis using the observed data set and others independent
  observational results. However, Whitmore \& Murphy \cite{Whit} have
  shown that long-range wavelength distortions can be confused and
  appreciated as if they were the reported dipolar variation of
  $\alpha$ \cite{murphy16}. Later, Murphy et al. \cite{murphy17}
  studied certain transitions that do not suffer from long-range
  distortions obtaining results consistent with the null
  variation.}. In Tables \ref{tablaMerMart}, \ref{tablaEvans} and
\ref{tablaMurphy} we show results obtained by different authors using
different methods and absorption lines of different chemical elements.
\renewcommand{\arraystretch}{1.7}
\begin{table}[H]
  \begin{center}
    \begin{tabular}{cccc}
      \hline\hline
      {\bf Method/Quasar}&{\bf Redshift}&{\bf $\frac{\Delta\alpha}{\alpha_0}\pm\sigma [10^{-5}]$}&{\bf Ref.}\\ \hline
      SIDAM & $1.15$ &$ 0.04\pm 0.15$ & \citep{lev2007}\\
      SIDAM & $1.15$ & $024\pm 0.38$ & \citep{lev2007}\\
      OH conjugated lines &$025$& $0.5\pm 1.26$& \citep{dar2004}\\
      Molecular and radio lines&$0.24$ &$ -0.10\pm0.22$&\citep{mur2005}\\
      Molecular and radio lines&$0.69$&$ -0.08\pm0.27$&\citep{mur2005}\\
      Molecular and radio lines&$ 0.765$& $<0.67$ & \citep{kane2005}\\
      OH conjugated and radio lines&$ 0.865$&$<0.67$&\citep{kane2005}\\
      3 sources& $1.08$ & $0.43\pm 0.34$ & \citep{martins25}\\
      HE0515$-$ 4414&$1.15$&$-0.01\pm0.18$&\citep{martins26}\\
      HE0515$-$ 4414&$1.15$&$0.05\pm0.24$&\citep{martins27}\\ 
      HE0001$-$2340&$1.58$&$-0.15\pm0.26$&\citep{martins8}\\
      HE1104$-$1805A&$1.66$&$-0.47\pm0.53$&\citep{martins25}\\
      HE2217$-$2818&$1.69$&$0.13\pm0.26$&\citep{martins8}\\
      HS1946$+$7658&$1.74$&$-0.79\pm0.26$&\citep{martins25}\\
      Q1101$-$264&$1.84$&$0.57\pm0.27$&\citep{martins25}\\\hline
    \end{tabular}
    \caption{Bounds for the time variation of $\alpha$ from quasar
      absorption systems.}
    \label{tablaMerMart}
  \end{center}
\end{table}
\renewcommand{\arraystretch}{1.7}
\begin{table}[H]
  \begin{center}
    \begin{tabular}{ccc}
      \hline
      {\bf Telescope} & {\bf Redshift} &{\bf $\frac{\Delta\alpha}{\alpha_0}\pm\sigma_{\rm stat}\pm\sigma_{\rm sys}[10^{-6}]$}\\\hline\hline
      Keck/HIRES&$1.143$&$0.20\pm13.63\pm3.97$\\
      VLT/UVES&$1.143$&$-8.80\pm5.60\pm4.36$\\
      Subaru/HDS&$1.143$&$-9.04\pm10.41\pm4.34$\\
      Keck/HIRES&$1.342$&$-2.77\pm13.71\pm3.16$\\
      VLT/UVES&$1.342$&$0.02\pm7.64\pm1.85$\\
      Subaru/HDS&$1.342$&$-1.29\pm24.04\pm6.04$\\
      Keck/HIRES&$1.802$&$-3.92\pm8.61\pm4.69$\\
      VLT/UVES&$1.802$&$-0.66\pm14.65\pm4.54$\\
      Subaru/HDS&$1.802$&$-11.20\pm7.83\pm2.44$\\ \hline
    \end{tabular}
    \caption{Bounds on the time variation of the fine structure
      constant of three different absorption systems from Quasar
      HS1549 + 1919 \cite{Evans2014}.}
    \label{tablaEvans}
  \end{center}
\end{table}
\renewcommand{\arraystretch}{1.7}
\begin{table}[H]
  \begin{center}
    \begin{tabular}{cccc}
      \hline
      {\bf Quasar} & {\bf Redshift} & {\bf Telescope} &{\bf $\frac{\Delta\alpha}{\alpha_0}\pm\sigma_{\rm stat}\pm\sigma_{\rm sys}[10^{-6}]$}\\ \hline\hline
      J0058+0041 & $1.072$ &  Keck & $-1.35\pm6.71 \pm2.51$\\ 
      J0058+0041 & $1.072$ &  VLT  & $17.07\pm9.00\pm2.41$\\ 
      PHL957 & $2.309$ & Keck & $ -0.65\pm6.46\pm2.26$\\ 
      PHL957 & $2.309$ & VLT & $-0.20 \pm12.44 \pm3.51$\\ 
      J0108-0037 & $1.371$ & VLT & $ -8.45\pm5.69\pm4.64$\\ 
      J0226-2857 & $1.023$ & VLT & $3.54\pm8.54 \pm2.38$\\ 
      J0841+0312 & $1.342$ & Keck & $3.05\pm3.30\pm2.13$\\ 
      J0841+0312 & $1.342$ & VLT & $5.67\pm4.19\pm2.16$\\ 
      J1029+1039 & $1.622$ & Keck & $-1.70\pm9.80\pm2.47$\\
      J1237+0106 & $1.305$ & Keck & $-4.54\pm8.08\pm3.13$\\ 
      Q1755+57 & $1.971$ & Keck & $4.72\pm4.18\pm2.16$\\ 
      Q2206-1958 & $1.921$ & VLT & $-4.65\pm6.01\pm2.24$\\ \hline
    \end{tabular}
    \caption{Bounds on the time variation of the fine structure
      constant from different quasars \cite{murphy17}.}
    \label{tablaMurphy}
  \end{center}
\end{table}

\subsection{Oklo constraint}
Oklo is a natural uranium-fusion nuclear reactor that operated
$1.8\times10^9$ years ago in Gabon, Africa. The operating conditions
of the reactor can be recovered from the analysis of nuclear and
geochemical data to calculate the thermal neutron capture cross
sections of several nuclear species including $^{149}$Sm, $^{151}$Eu
and $^{155}$Gd. From these cross sections it is possible to estimate
the value of the resonance energy of the fundamental level at the time
of the reaction. A variation of this energy (from the time of the
reaction to the present) would reflect a variation in time of $\alpha$
\cite{Damour,Fuji,sisterna,Lamo04}. The time variation of the fine
structure constant obtained with this natural reactor is \cite{Lamo04}
\begin{eqnarray}
  \frac{\Delta\alpha}{\alpha_0}&=&\left(45\pm15\right) \times 10^{-9} .
\end{eqnarray}

\subsection{Long-lived $\beta$ decayers constraints}
Thanks to laboratory measurements and/or by comparison with the age of
meteorites, it is possible to determine the half-life of long-lived
$\beta$ decayers. Sisterna \& Vucetich \cite{sisterna2} have analyzed
the dependence between the half-life of these decayers and the
fundamental constants $\alpha$, $\Lambda_{\rm QCD}$ and $G_F$ so, any
shift in the half life may be generated by a variation in these
constants at the age of the meteorites and their value now. According
to these authors, there is a linear relationship between the shift in
the half-life of the $\beta$ decayer of $^{187}$Re from the solar
system formation up today and the variation of $\alpha$. The
transition analyzed was Re $\rightarrow$Os and it is assumed that this
event has occurred $4.6\times10^{9}$ years ago. According to
Ref. \cite{sisterna2}, the time variation of the fine structure
constant is
\begin{equation}
  \frac{\Delta\alpha}{\alpha_0}=\left(-7.4\pm7.4\right)\times 10^{-7}.
\end{equation}

\subsection{Atomic clocks constraints}
The time variation of the fine structure constant can be measured by
comparing frequencies of atomic clocks with different atomic numbers
during time intervals that range from one hundred days to two
years. This is possible thanks to the development of very stable
frequency oscillators based on hyperfine transitions. These hyperfine
levels are determined by the interaction of the nuclear magnetic
moment with the magnetic moment of a valence electron.  It is possible
to restrict the variation of $\alpha$ by comparing the frequencies of
two different atoms because the relativistic contribution to the
splitting of the hyperfine levels $\Delta$ grows with the atomic
number as $\Delta \propto \left(Z\alpha\right)^2$. Table \ref{Relojes}
shows the bounds for the time variation of the fine structure constant
presented in the literature.  \renewcommand{\arraystretch}{1.7}
\begin{table}[H]
  \begin{center}
    \begin{tabular}{ccc}
      \hline
      {\bf Frequencies}&{\bf $\frac{\dot\alpha}{\alpha_0}$ [$10^{-15}$ yr$^{-1}$]}& {\bf Ref.}\\     \hline \hline
      Hg$^{+}$ y H maser & $0.0 \pm 14.0$ & \cite{Prestage1995}\\
      C$_{\rm s}$ y R$_{\rm b}$& $8.4\pm13.8$ & \cite{sortais}\\
      C$_{\rm s}$ y R$_{\rm b}$&$-0.2\pm 8.0$& \cite{marion}\\
      H$_{\rm g}$ y C$_{\rm s}$&$5.7\pm11.2$&\cite{fischer}\\
      Y$_{\rm b}$ y C$_{\rm s}$&$-1.6\pm5.9$&\cite{peik}\\
      C$_{\rm s}$&$0.33\pm0.3$&\cite{li}\\
      H$_{\rm g}$&$0.053\pm0.079$&\cite{li}\\
      H$_{\rm g}^{+}$&$-0.016\pm0.023$&\cite{LoriniL} \\
      \hline
    \end{tabular}
    \caption{Bounds on $\dot{\alpha}/\alpha_0$ from different atomic
      clocks.}
    \label{Relojes}
  \end{center}
\end{table}

\section{Results}
\label{sec:results}
In this section we analyze and discuss the predictions that emerge
from the {\it improved Bekenstein model in the semiclassical regime}
and compare it with experimental and observational data in order to
verify the validity of the theory. We analyze separately the spatial
and temporal aspects of the theory.

\subsection{WEP experiments for testing the spatial variation of
  $\alpha$}
\label{sec:EotWash}
As we have explained in Section \ref{KVAP}, the E\"otv\"os parameter
can be obtained from the difference of the acceleration of two
different bodies of different composition. In order to determine
and/or limit the validity of the theory, we use the data from
E\"ot\"os-type experiments presented in Table \ref{tab:EotRes}.

Using Eq. (\ref{acbsc}), the theoretical prediction of the E\"otv\"os
parameter for the semiclassical Bekenstein model is
\begin{eqnarray}
  \eta(A, B) = \frac{a_A - a_B}{g}& =& 4 \left(\frac{\ell_B}{\ell_P} \right)^2 \zeta^S_m (\zeta^{A}_m - \zeta^{B}_m)  -24 \left(\frac{\ell_B}{\ell_P} \right)^4 \left(\zeta^S_m\right)^2 (\zeta^{A}_m - \zeta^{B}_m) \frac{\phi_N(R_{s})}{c^2} \label{acelpsi1} \nonumber \\
                                  &=& \widetilde{A}  \left(\frac{\ell_B}{\ell_P}\right)^2 + \widetilde{B}  \left(\frac{\ell_B}{\ell_P}\right)^4, \label{eta_coef}
\end{eqnarray}
where $\widetilde{A}$ and $\widetilde{B}$ are constant that depend on
the composition of the bodies and do not depend on the theoretical
parameter of the model. Note that the Eq. (\ref{eq:Bek:eta:expr}) can
be written as
$\eta=\widetilde{A} \left(\frac{\ell_B}{\ell_P}\right)^2$. We perform
a $\chi^2$-test to set constrains on the free parameter of the theory,
namely $\ell_B/\ell_P$, by the comparison of our theoretical results
with the observational data. In Table \ref{valorcoef} we show our
results and we also compared, in the same table, the results obtained
using the approximation of Ref. \cite{LK11a} or the
Eq. (\ref{eq:Bek:eta:expr}). The differences between this new limit and
the one presented in Eq. (\ref{eq:Res1:lBlP2}) are due to the
improvement in the estimation of the magnetic energy (see Appendix
\ref{NME}).  \renewcommand{\arraystretch}{1.7}
\begin{table}[H]
  \begin{center}
    \begin{tabular}{ccc}
      \hline
      {\bf Approximation} & {\bf $\left(\frac{\ell_B}{\ell_P}\right)^2\pm\sigma$}& {${\chi}^2_{\nu}$} \\ \hline\hline
      $\eta= \widetilde{A}\left(\frac{\ell_B}{\ell_P}\right)^2+\widetilde{B}\left(\frac{\ell_B}{\ell_P}\right)^4 $& $(7.9\pm 70.7)\times10^{-5}$& $0.329$  \\  
      $\eta= \widetilde{A}\left(\frac{\ell_B}{\ell_P}\right)^2$ & $(7.9\pm 70.7)\times10^{-5}$& $0.329$ \\\hline
    \end{tabular}
    \caption{Estimated value for the free parameter in two different
      theoretical expressions. $\chi^2_{\nu}$ stand for the normalized
      $\chi^2$.}
    \label{valorcoef}
  \end{center}
\end{table}   

As one can noticed, the results for both cases are the same, since the
contribution from the ``next order'' (first order) in the series
expansion of the field $\psi$ is very small and can be
neglected. Consequently, it is valid the use of the zero order
approximation with the change on the estimation of the magnetic
energy.

Since at $3\sigma$-level, the upper bound of the Bekenstein length
scale, for both cases, is $\ell_B < 0.05 \ell_P$, it seems naturally
to discard the theory.

In December 2017, the MICROSCOPE collaboration published an estimation of $\eta\sim10^{-15}$ consistent with zero \cite{Tou17}. If this result is used, the $\frac{\ell_B}{\ell_P}$ bounds would be even smaller.

\subsection{Constrains from the time variation of $\alpha$}
\label{sec:time}
In the Section \ref{magtemporal} we have computed the theoretical
predictions for the variation of the fine structure constant in
different stages of the Universe. Eqs. (\ref{var_al_ut}),
(\ref{var_al_um}) and (\ref{var_al_ua}) can be written as
\begin{subequations}
  \begin{eqnarray}
    \frac{\Delta\alpha}{\alpha} &=& \breve{A} \left(\frac{\ell_B}{\ell_P}\right)^2+ \breve{B} H_0t_c \left(\frac{\ell_B}{\ell_P}\right)^2 , \label{UniTempyMedio} \\
    \frac{\dot\alpha}{H_0\alpha} &=& \breve{C} \left(\frac{\ell_B}{\ell_P}\right)^2+ \breve{C} H_0t_c \left(\frac{\ell_B}{\ell_P}\right)^2,\label{UniActual}
  \end{eqnarray}
\end{subequations}
where the constants $\breve{A}$, $\breve{B}$ and $\breve{C}$ depend
only on the cosmological parameters. The constraints on the free
parameters of the theory, that is
$\left(\frac{\ell_B}{\ell_P}\right)^2$ and
$H_0t_c \left(\frac{\ell_B}{\ell_P}\right)^2$, are obtained by a least
squares linear regression using the observational data described in
Section \ref{sec:data}. We have performed a discrimination into
different groups of the available data, according to when the physical
phenomena took place, that is: i) early Universe (CMB and BBN bounds);
ii) middle-age Universe (Quasars, Long-live $\beta$ decayers and Oklo
reactor bounds); and iii) present Universe (bounds from atomic
clocks). As we have mentioned before, we use three different sets of
cosmological parameters.

In Table \ref{times} we present our results for the constraints on the
free parameter for the different models (or cosmological parameters)
described in Section \ref{sec:BBN}. The value for the variation of the
fine structure constant during BBN used in the lineal regression
correspond to the obtained in Section \ref{sec:BBN} when the lithium
data was discarded in the statistical test.
\renewcommand{\arraystretch}{1.7}
\begin{table}
  \begin{center}
    \begin{tabular}{ccccc} \hline {\bf Model} &{\bf Data}& {\bf
        $\left(\frac{\ell_B}{\ell_P}\right)^2 \pm \sigma$} & {\bf
        $ H_0 t_c \left(\frac{\ell_B}{\ell_P}\right)^2 \pm \sigma$}&
      {${\chi}^2_{\nu}$} \\ \hline\hline
                                              &All data &$-2.54\pm 0.84$& $(-1.82\pm1.76)\times10^{-6}$& $0.75$\\
      Model I & Without early Universe& $-2.54\pm 0.84$&$0.053\pm1.025$&$0.56$ \\
                                              &  Without middle-age Universe&$-1.05\pm2.68$&$(-1.82\pm 1.76)\times10^{-6}$&$1.29$ \\
                                              & Without present
                                                Universe&$
                                                          -2.70\pm0.88$&$(-1.82\pm1.76)\times10^{-6}
                                                                         $&$0.85$
      \\ \hline
     
                                              &All data &$-3.55\pm 1.22$& $(-2.58\pm15.50)\times10^{-7}$& $0.57$\\
      Model II&Without early Universe& $-3.55\pm 1.22$&$0.072\pm1.373$&$0.57$ \\
                                              &  Without middle-age Universe&$-1.10\pm2.76$&$(-2.55\pm 15.50)\times10^{-7}$&$0.36$ \\
                                              & Without present Universe&$ -4.14\pm1.35$&$(-2.57\pm15.51)\times10^{-7} $&$0.61$ \\ \hline
     
                                              & All data &$-4.20\pm 1.54$& $(3.66\pm1.72)\times10^{-6}$& $0.57$\\
      Model III & Without early Universe& $-4.20\pm 1.54$&$0.084\pm1.864$&$0.60$ \\
                                              &  Without middle-age Universe&$-1.11\pm2.72$&$(3.66\pm 1.72)\times10^{-6}$&$0.30$ \\     
                                              & Without present Universe&$ -5.69\pm1.87$&$(3.66\pm1.72)\times10^{-6} $&$0.60$ \\ \hline   
    \end{tabular}
    \caption{Best fits of the free parameters of the model for the different values of the cosmological parameters (see Tables \ref{BBN} and \ref{tablaUniTemp}).}
    \label{times}
  \end{center}
\end{table}
It can be noticed that in the three models, in the cases where the
data corresponding to the early Universe are subtracted, the value of
$ H_0 t_c \left(\frac{\ell_B}{\ell_P}\right)^2$ and its uncertainty
increase considerably. In addition, it can be seen that at one
standard deviation, Bekenstein length parameter $\ell_B$ becomes
imaginary for all the cases with the exception of the analysis
performed when the middle-age Universe data is excluded.

An important difference arises from the results for the different
groups of cosmological parameters. The estimates for
$ H_0 t_c \left(\frac{\ell_B}{\ell_P}\right)^2$ are negative when we
use the results of the cosmological parameters of Models I and II,
whereas the estimates using the parameters of Model III are positive
(except the case where BBN and CMB data are excluded).

Since several authors consider the integration constant $t_c$ null, we
have also performed an estimation of parameter
$\left(\frac{\ell_B}{\ell_P}\right)^2$ for each group of cosmological
parameters $t_c=0$, that is
$
\frac{\Delta\alpha}{\alpha_0}=\breve{A}\left(\frac{\ell_B}{\ell_P}\right)^2$
and
$\frac{\dot{\alpha}}{H_0\alpha_0}=\breve{C}\left(\frac{\ell_B}{\ell_P}\right)^2$. The
results presented in Table \ref{timestc0} are similar to the previous
ones, however the uncertainties are qualitatively greater. This
generates that only at 1$\sigma$ level the model does not fulfil the
hypothesis imposed on $\ell_B$, either because the estimates give
imaginary values or smaller values than the Planck length. On the
other hand, in the cases where the middle-aged Universe data are
eliminated, the model can not be discarded.

\renewcommand{\arraystretch}{1.7}
\begin{table}
  \begin{center}
    \begin{tabular}{cccc}\hline {\bf Model}&{\bf Data} & {\bf
                                                         $\left(\frac{\ell_B}{\ell_P}\right)^2 \pm \sigma$} & {\bf
                                                                                                              $\chi^2_{\nu}$} \\ \hline\hline
                                           &All data&$-2.54\pm2.22 $& $0.76 $\\
      Model I& Without early Universe& $-2.54\pm 2.19$&$0.55$ \\
                                           & Without middle-age Universe&$-1.05\pm4.95$&$1.27$ \\
                                           & Without present Universe&$ -2.69\pm2.03$&$0.86$ \\ \hline
        
                                           & All data &$-2.99\pm2.71 $& $0.56 $\\
      Model II& Without early Universe& $-2.99\pm 2.67$&$0.56$ \\
                                           &  Without middle-age Universe&$-1.10\pm5.09$&$0.33$ \\
                                           &  Without present Universe&$ -2.99\pm2.19$&$0.61$ \\\hline  
     
                                           &All data&$-2.99\pm3.05 $& $0.67 $\\
      Model III& Without early Universe& $-2.99\pm 3.01$&$0.60$ \\
                                           & Without middle-age Universe&$-1.10\pm5.02$&$0.66$ \\
                                           &  Without present Universe&$ -2.99\pm2.42$&$0.74$ \\ \hline     
    \end{tabular}
    \caption{Best fits of the free parameters of the model for the different values of the cosmological parameters and $t_c=0$ (see Tables \ref{BBN} and \ref{tablaUniTemp}).}
    \label{timestc0}
  \end{center}
\end{table}   

\section{Summary and Conclusions}
\label{sec:conclusions}

In this work we have analyzed the magnetic contributions in the
semiclassical Bekenstein model for the variation of the fine structure
constant and show that the magnetic field produces an anomalous term
in the acceleration. Based on previous works of
Refs. \cite{LK11a,LK12}, we have solved the equation of motion the
scalar field $\psi$ to first order (in the mentioned publications this
calculation was made at zero order). We have also improved the
estimation of nuclear magnetic energy, according to
Refs. \cite{Levi60,BF75,Gor}. To set constrains on the Bekenstein's
theory parameter, $\ell_B$, we have analyzed not only the spatial
contribution of the equation of motion of the field related to the
fine structure constant, but also its time variation. To perform the
last computation we have considered the expansion rate of the Universe
in different evolutionary stages.


We have examined possible violations of the 
Weak Equivalence Principle due to the incorporation
of the scalar field. In order to constrain the Bekenstein length, we
have compared our theoretical results with the experimental data from
E\"otv\"os-type experiments. We conclude that the first term in the
series of the weak field approximation is the one that contributes the
most and the following ones can be neglected. Our
  results are of the same order as those obtained in
  Ref. \cite{LK11a}, and the small differences between them are due
  only to the change in the estimation of the magnetic energy.
Consequently, it can be deduced that the semiclassical model
for the spatial variation would be discarded.

We have also analyzed the time variation of the fine structure
constant from the semiclassical model of Bekenstein in an expanding
Universe. To determine the free parameters of the theory,
$\left(\frac{\ell_B}{\ell_P}\right)^2$ and
$\left(\frac{\ell_B}{\ell_P}\right)^2H_0 t_c $, we have made a
comparison between the observational data for different times with our
theoretical predictions. At $3\sigma$ level, we found
\begin{subequations}
  \begin{equation}
    \left(\frac{\ell_B}{\ell_P}\right)^2 < -0.02, \qquad
    \left(\frac{\ell_B}{\ell_P}\right)^2 < 0.11, \qquad
    \left(\frac{\ell_B}{\ell_P}\right)^2 < 0.42,
  \end{equation} 
  \begin{equation}
    \frac{\ell_B}{\ell_P} < 0.14\, \dot{\imath}, \qquad
    \frac{\ell_B}{\ell_P} < 0.33, \qquad
    \frac{\ell_B}{\ell_P} < 0.65,
  \end{equation}
\end{subequations}
for the different cosmological parameters used, that is the first
constrain correspond to the Model I, the second to the Model II and
the last one to Model III. Note that, for the first expression, Bekenstein length turns out to be a non-real number. These results rule out the theory.

The removal or not of the early Universe data (that is BBN and CMB
constrains on the variation of the fine structure constant) does not
affect the results. The same constrains are obtained if the data
removed are those from the present Universe (that is bound obtained
from atomic clocks). However, in the case where the data of the
middle-age Universe are not taken into account in the fit, the
$3\sigma$ level upper bound does not allow to discard the theory since
$\ell_B\sim\ell_P$.

The $3\sigma$ level upper bound results for the parameter
$\left(\frac{\ell_B}{\ell_P}\right)^2H_0 t_c$ are (in the same order
as before)
\begin{equation}
  \left(\frac{\ell_B}{\ell_P}\right)^2H_0 t_c < 3\times10^{-6},\qquad
  \left(\frac{\ell_B}{\ell_P}\right)^2H_0 t_c < 4\times10^{-6},\qquad
  \left(\frac{\ell_B}{\ell_P}\right)^2H_0 t_c < 8\times10^{-6}.
\end{equation}

We obtained the same results when we excluded data from the
middle-aged Universe and/or the present Universe. On the contrary,
when the early Universe data are not taken into account the bounds are
much less restrictive, that is
\begin{equation}
  \left(\frac{\ell_B}{\ell_P}\right)^2H_0t_c < 3,\qquad
  \left(\frac{\ell_B}{\ell_P}\right)^2H_0t_c < 4,\qquad 
  \left(\frac{\ell_B}{\ell_P}\right)^2H_0t_c < 6.
\end{equation}\\

  Comparing our results with those obtained in
  Ref. \cite{mosquera08} where possible time variations of $\alpha$
  due to the Coulomb contributions are analyzed, it is observed that
  for the parameter $\left(\frac{\ell_B}{\ell_P}\right)^2H_0t_c$ both
  set of bounds are of the same order, while for parameter the
  $\left(\frac{\ell_B}{\ell_P}\right)^2$ their bounds are more tighten
  (around an order of magnitude or two smaller).

Finally, the results for the Bekenstein length when $t_c=0$ indicate
that only at one standard deviation the theory can be discarded in all
cases for the three models except when the data from the middle-aged
Universe are not considered in the statistical
analysis. Contrasting these results with those
  obtained from the analyses of the planet's thermal flow (data from
  the late Universe) \cite{KVMS,LK12}, their 3-$\sigma$ upper bounds
  are of an order of magnitude more strict than ours using data of the
  whole evolution of the Universe.\\

It is important to note that almost all the mean values obtained for the parameters $\left(\frac{\ell_B}{\ell_P}\right)^2$  and $\left(\frac{\ell_B}{\ell_P}\right)^2H_0t_c$  from the analysis of the time variation of $\alpha$ are negative, producing an incongruity in the semiclassical model since Bekenstein length must be real, positive and greater than Planck length. This fact gives us a hint of the serious problems that the theory faces although the given bulging uncertainties.

In conclusion, it has been shown that even taking as valid
Bekenstein's statement (about the cancellation in the source term
between the contributions of electric field and the mass) in a
semiclassical model, the magnetic contributions produce observable
anomalous accelerations and time variations of the fine structure
constant which are not compatible with the observable data.
 
\section*{Acknowledgments}
L.K. is supported by PIP 11220120100504 CONICET, and with H.V. by
grant G140 UNLP.  M.E.M. is supported by a grant (PIP-282) of the
National Research Council of Argentina (CONICET) and by a research-grant of the National Agency for the Promotion  of Science and Technology (ANPCYT) of Argentina. L.K. and M.E.M. are
members of the Scientific Research Career of the CONICET.

\appendix
\section{The integrated strength function in the nuclear magnetic
  energy}
\label{NME}
As mentioned in Section (\ref{KVAP}), magnetic energy density is
mainly located close to the atomic nuclei, reason for
  studying its contributions are studied.  Another possible
  source of magnetic field are the quantum fluctuations of the
``number of particles''. The main contribution to the magnetic energy
under the {\it semi-classical} scheme comes from the dipolar nuclear
oscillations with T=1 (isospin). The protons and neutrons oscillate in
anti-phase, generating a variable dipole moment  and, therefore a
variable current. For a detailed description of  the computation
  of magnetic energy by relating the current to the dipole transition
matrix elements with the Thomas-Reiche-Kuhn sum rule see Ref.
\cite{LK11a}. This sum rule is saturated (approximately) by the giant
dipole resonance.

In this work we have a special interest in improving the 
  computation of the integrated strength function,  that is
$\int\sigma(E) dE$ of Eq. (\ref{eq:Bek:Mag:Ener}) for different
  nuclei. For most cases the rule of Thomas-Reiche-Kuhn \cite{Levi60}
gives a good estimate
\begin{equation}
  \label{eq:TRK:Sum:Rule}
  \sigma_T=\int\sigma(E)dE = (1+x)\frac{2\pi^2e^2\hbar}{mc} \frac{NZ}{A} \simeq  60\frac{NZ}{A} {\rm MeV}{\rm mb},
\end{equation}
where $x\sim0.2$ takes into account the interchange and rate of
dependence of nuclear interactions, $N$, $Z$ and $A$ are the
neutron, proton and mass numbers ($A = N + Z$) of the
element with mass $m$.  However, for light nuclei such as hydrogen ($N=0$), helium ($N=2$) or  beryllium ($N=5$ for $^9$Be), this estimation fails. For this reason, we
apply, in these  particular cases the
results obtained  in Ref. \cite{BF75}, where the effective section
  of $^4{\rm He}$ was measured
  ($\sigma_T\approx 7.94 {\rm MeV}{\rm mb}$); and in Ref. \cite{Gor},
  where the proton resonances were analyzed with the
    result 
    $\sigma_T\approx 46 {\rm MeV} {\rm mb}$. On the other hand, from the photodisintegration of $^9$Be through the 1/2+ state near neutron threshold analysis \cite{Utsu}, the best estimate obtained is $\sigma_T\approx 13.3 {\rm MeV} {\rm mb}$.


\end{document}